# Effective Local Permittivity Model for Non-Local Wire Media

Alexander B. Yakovlev, *Senior Member*, *IEEE*, Mário G. Silveirinha, *Fellow*, *IEEE*, and George W. Hanson, *Fellow*, *IEEE*

*Abstract*—A local permittivity model is proposed to accurately characterize spatial dispersion in non-local wire-medium (WM) structures with arbitrary terminations. A closed-form expression for the local thickness-dependent permittivity is derived for a general case of a bounded WM with lumped impedance insertions and terminated with impedance surfaces, which takes into account the effects of spatial dispersion and loads/terminations in the averaged sense per length of the wire medium. The proposed approach results in a local model formalism and accurately predicts the response of WM structures for near-field and far-field excitation. It is also shown that a traditional transmission network and circuit model can be effectively used to quantify the interaction of propagating and evanescent waves with WM structures. In addition, the derived analytical expression for the local thickness-dependent permittivity has been used in the full-wave numerical solver (CST Microwave Studio) demonstrating a drastic reduction in the computation time and memory in the solution of near-field and far-field problems involving wire media.

## I. Introduction

SPATIAL DISPERSION (SD) of continuous materials (or homogenized materials with effective material parameters) characterizes the dependence of constitutive parameters on wavevector, such that the constitutive relations between the macroscopic fields and electric/magnetic dipole moment are non-local [1], [2]. Non-local effects generate extra propagating or evanescent waves (extraordinary waves in regard to local materials). The wave-matter interaction in bounded non-local materials is typically modeled by expanding the field in terms of the waves of bounded domains (as if they were infinite), and then by imposing the usual boundary conditions as well as additional boundary conditions (ABCs) at material interfaces to account for the extra waves in non-local materials [3], [4].

However, even though the field decomposition using the waves of an infinite non-local material results in a physical material response, the constitutive relations in the spatial transform-domain for a bounded non-local material are no longer represented by a convolution integral, which holds for a translationally invariant media (such as infinite media), and the material response must take into account the material boundary [3], [4].

In this paper, we focus on the analysis of electromagnetic interactions with bounded WM structures. It is already well known that wire media exhibits strong spatial dispersion at microwave frequencies even for very long wavelengths [5], [6]. The role of SD effects has been addressed [7]-[10], and it has been shown that the non-local homogenization formalism is essential in the solution of scattering, radiation, and excitation electromagnetic problems involving wire media and WM-type structures [8], [9], [11]-[24]. In this regard, the methodology proposed in [25] is of particular interest, wherein the non-local susceptibility (non-local permittivity) of the bounded non-local homogenized WM in the spatial transform-domain is given by the Green's function for the same geometric region subject to ABCs at the wire-end terminations [26]-[29]. Based on the non-local formalism for bounded WM proposed in [25], a local thickness-dependent permittivity has been derived in closed form for a grounded WM and symmetric WM terminated with impedance surfaces at the wire connections [30]. It was shown that the local permittivity that accounts for spatial dispersion must depend on the thickness of the WM slab and the termination of the wires. In this regard, it is worth to point out Ref. [31], which discusses the dependence of effective material parameters on geometric metamaterial parameters. Also, in recent work [32] it is shown that the spatial non-locality in metals can be effectively modeled by a composite material comprising a thin local dielectric layer on top of a local metal, such that the non-local effects are captured in a deeply subwavelength effective dielectric layer. In general, this is not the case with WM, wherein the non-local effects are distributed throughout the entire WM due to the presence of two extraordinary waves (transverse magnetic (TM), which is evanescent below the plasma frequency, and transverse electromagnetic (TEM), which propagates in WM as in an uniaxial material with extreme anisotropy). However, for a thin (subwavelength) WM slab the local thickness-dependent permittivity can be regarded as an effective local permittivity which captures the non-local effects similar to the thin local dielectric layer in metals studied in [32].

In the study presented here we generalize the concepts developed in [25], [30], [33], and [34] to the case of WM slab with lumped impedance insertions (including the case of PIN diodes at the wire connections) and terminated with different impedance surfaces at the top and bottom WM interfaces (as an asymmetric configuration). The local thickness-dependent permittivity is derived in closed form, and takes into account spatial dispersion and the effect of the loads/terminations in the averaged sense per thickness of the WM slab. The local thickness-dependent permittivity for an asymmetric configuration with lumped impedance insertions and terminated with arbitrary impedance surfaces first appeared in [35], and here the theory is verified with full-wave results, and several other geometries/examples have been added.

This paper is organized as follows. In Section II, we introduce the formalism of the non-local susceptibility (non-local permittivity) in the spatial domain for a WM slab with lumped loads/impedance surfaces terminations, and derive a closed-form expression for the local thickness-dependent permittivity. Also, a transmission network approach and circuit model are discussed here. In Section III, various numerical examples in the local framework are demonstrated and the results are compared with the non-local solution and full-wave numerical simulations. Section IV summarizes the main results and conclusions. Also, the paper is accompanied by an appendix with the analytical details of the Green's function problem. A time dependence of the form $e^{j\omega t}$ is assumed and suppressed.

II. LOCAL THICKNESS-DEPENDENT PERMITTIVITY OF WM

Consider a WM structure with the geometry shown in Fig. 1. The structure is comprised by a 2D lattice of $z-$directed thin metallic wires of thickness $L$ connected through lumped impedance insertions to impedance surfaces at the wire-end interfaces at $z=0$ and $z=-L$. For simplicity, we assume air semi-infinite regions external to the structure. The period of wires is $a$, the radius of wires is $r_0$, and the relative permittivity of the host medium is $\varepsilon_h$. The impedance surfaces are characterized by the surface admittance, $\bar{\mathbf{Y}}_{g1,2} = (\hat{\mathbf{x}}\hat{\mathbf{x}} + \hat{\mathbf{y}}\hat{\mathbf{y}})Y_{g1,2}$, with the closed-form expressions for periodic printed (capacitive) or slotted (inductive) subwavelength grids presented in [36]. The lumped impedance insertions are given in terms of the effective load impedance, $Z_{\text{load eff1,2}}$ [23], [28], [37].

According to [25], the non-local response for the geometry in Fig. 1 that accounts for a material boundary in the spatial transform-domain satisfies the following system of equations written from a macroscopic perspective for a homogenized WM,

$$\nabla \times \mathbf{E} = -j\omega\mu\mathbf{H} \quad (1)$$

$$\nabla \times \mathbf{H} = \left(\bar{\mathbf{Y}}_{g1}\delta(z) + \bar{\mathbf{Y}}_{g2}\delta(z+L)\right)\cdot\mathbf{E} + j\omega P_z^{\text{cond}}\hat{\mathbf{z}} + j\omega\varepsilon_0\varepsilon_h\mathbf{E} \quad (2)$$

$$\left(k_h^2 + \frac{\partial^2}{\partial z^2}\right)P_z^{\text{cond}} = -k_p^2\varepsilon_0\varepsilon_h E_z \quad (3)$$

where $P_z^{\text{cond}}$ is the conductive polarization due to the $z-$directed WM,

$$P_z^{\text{cond}}(z) = \varepsilon_0 \int_{-L}^{0} \chi^{\text{cond}}(z,z')E_z(z')dz'. \quad (4)$$

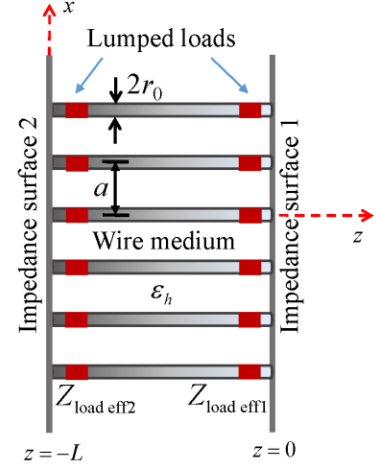

Fig. 1. Geometry of a 2D WM structure with the wires connected to impedance surfaces through lumped impedance insertions.

Here, the susceptibility $\chi^{\text{cond}}$ determines the non-local response for the uniaxial WM that accounts for the material interface. It is proportional to the Green's function associated with (3) subject to the ABCs at the interfaces, such that $\chi^{\text{cond}}(z,z') = \varepsilon_h k_p^2 g(z,z')$. The Green's function problem for the geometry in Fig. 1 is presented in the appendix. In (1)-(3), $k_h = k_0\sqrt{\varepsilon_h}$ is the wavenumber of the host medium, and $k_p$ is the plasma wavenumber defined by $(k_p a)^2 \approx 2\pi / (\ln(a/(2\pi r_0)) + 0.5275)$ [6].

The non-local permittivity for WM can be described in terms of the Green's function, $g(z,z')$, which takes into account the material spatial dispersion and the effect of the lumped loads and the material boundaries (Fig. 1),

$$\varepsilon_{\text{nonloc}}(z,z') = \varepsilon_h k_p^2 g(z,z') + \varepsilon_h \delta(z-z'). \quad (5)$$

With the known Green's function (see Appendix) the non-local permittivity (5) can be approximated by a local response, $\varepsilon_{\text{nonloc}}(z,z') \approx \varepsilon_{\text{loc}}\delta(z-z')$, where $\varepsilon_{\text{loc}}$ is the *local thickness-dependent permittivity* [30] for a WM structure of thickness $L$ with the wires connected to impedance surfaces through lumped loads (Fig. 1) resulting in,

$$\varepsilon_{\text{loc}} = \frac{1}{L}\int_{-L}^{0}\int_{-L}^{0}\varepsilon_{\text{nonloc}}(z,z')dzdz'$$

$$= \frac{\varepsilon_h k_p^2}{L}\int_{-L}^{0}\int_{-L}^{0}g(z,z')dzdz' + \varepsilon_h.$$

(6)

Performing the double integral (6) with the Green's function (19), (20), a closed-form expression of the local thickness-dependent permittivity for a general case presented in Fig. 1 can be obtained,

$$\varepsilon_{\text{loc}} = \varepsilon_h\left(1 - \frac{k_p^2}{k_h^2}\right) + \varepsilon_h\frac{k_p^2}{Lk_h^3}\frac{2 - 2\cos(k_h L) + k_h(\alpha_1+\alpha_2)\sin(k_h L)}{(1-k_h^2\alpha_1\alpha_2)\sin(k_h L) + k_h(\alpha_1+\alpha_2)\cos(k_h L)}$$

(7)

where $\alpha_1$ and $\alpha_2$ are defined in the appendix for different cases of wire-end terminations. The first term in (7) is the usual local permittivity of a bulk WM (Drude permittivity), and the second thickness-dependent term takes into account spatial dispersion in the WM in the averaged sense per thickness $L$ and the effect of the loads/boundaries and the interaction between the boundaries. In a few limiting cases the obtained expression (7) simplifies to those presented in [30], such that for a grounded WM slab with impedance surface at $z=0$ ($\alpha_1 = \alpha$, $\alpha_2 \to \infty$) we obtain the expression (16) in [30], and for a grounded WM slab with open-end wires (no impedance surface at $z=0$ ($\alpha_1 = 0$, $\alpha_2 \to \infty$)) (7) results in (17) in [30]. Also, it can be verified that the limiting cases to symmetric structures (two-sided WM slab with identical impedance surfaces, $\alpha_1 = \alpha_2 = \alpha$, and the WM slab with open-end wires, $\alpha_1 = \alpha_2 = 0$) are also satisfied resulting in (16) and (17) in [30], respectively, by changing $L$ to $L/2$ in the obtained expressions.

The closed-form expression (7) for a local thickness-dependent permittivity can be used in the analysis of various near-field and far-field electromagnetic problems formulated on a local framework,

$$\nabla \times \mathbf{H} = j\omega\varepsilon_0 \bar{\bar{\varepsilon}}_{\text{total}}(z)\cdot \mathbf{E}(z) \quad (8)$$

where

$$\bar{\bar{\varepsilon}}_{\text{total}}(z) = \frac{1}{j\omega\varepsilon_0}\left(\bar{\bar{\mathbf{Y}}}_{g1}\delta(z) + \bar{\bar{\mathbf{Y}}}_{g2}\delta(z+L)\right) + \varepsilon_{\text{loc}}\hat{\mathbf{z}}\hat{\mathbf{z}} + \varepsilon_h\left(\hat{\mathbf{x}}\hat{\mathbf{x}} + \hat{\mathbf{y}}\hat{\mathbf{y}}\right).$$

(9)

The scattering problem for the geometry shown in Fig. 1 with the WM slab (including the lumped loads) replaced by the local uniaxial anisotropic material shown in Fig. 2 with the host permittivity $\varepsilon_h$ in the $x-$ and $y-$directions and the thickness-dependent permittivity (7) in the $z-$direction can be formulated in a local framework. We assume the TM-polarized plane wave is obliquely incident on the structure from the air region with $z>0$. The plane wave is characterized by $H_y$, $E_x$, and $E_z$ components, and the magnetic field in the air regions and in the WM slab can be expressed as follows,

$$H_y = \begin{cases} e^{\gamma_0 z} - \rho e^{-\gamma_0 z} & z > 0 \\ A^+ e^{\gamma_{\text{loc}} z} + A^- e^{-\gamma_{\text{loc}} z} & -L \leq z \leq 0 \\ Te^{\gamma_0(z+L)} & z < -L \end{cases} \quad (10)$$

where $\rho$ and $T$ are the reflection and transmission coefficients, respectively, $\gamma_0 = \sqrt{k_x^2 - k_0^2}$, $\gamma_{\text{loc}} = \sqrt{\frac{\varepsilon_h k_x^2}{\varepsilon_{\text{loc}}} - k_h^2}$ [30], and $k_x$ is the $x-$component of the wavevector $\mathbf{k} = (k_x, 0, k_z)$.

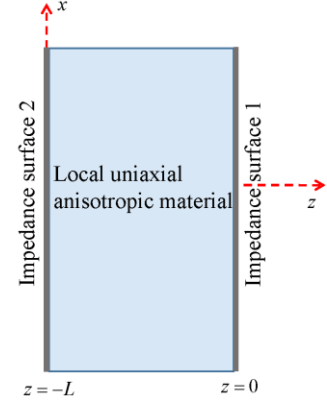

Fig. 2. A local uniaxial anisotropic material with impedance surfaces at the interfaces as an equivalent geometry to the WM structure presented in Fig. 1.

By imposing the usual boundary conditions for tangential electric-field and magnetic-field components at the interfaces at $z=0$ and $z=-L$ the following system of linear equations is obtained,

$$\begin{pmatrix} \varepsilon_h\gamma_0 & -\gamma_{\text{loc}} & \gamma_{\text{loc}} & 0 \\ 1+\frac{\gamma_0}{j\omega\varepsilon_0}Y_{g1} & 1 & 1 & 0 \\ 0 & \gamma_{\text{loc}}e^{-\gamma_{\text{loc}}} & -\gamma_{\text{loc}}e^{\gamma_{\text{loc}}} & -\varepsilon_h\gamma_0 \\ 0 & e^{-\gamma_{\text{loc}}} & e^{\gamma_{\text{loc}}} & -1-\frac{\gamma_0}{j\omega\varepsilon_0}Y_{g2} \end{pmatrix} \quad (11)$$

$$\times \begin{pmatrix} \rho \\ A^+ \\ A^- \\ T \end{pmatrix} = \begin{pmatrix} -\varepsilon_h\gamma_0 \\ 1-\frac{\gamma_0}{j\omega\varepsilon_0}Y_{g1} \\ 0 \\ 0 \end{pmatrix}$$

which can be solved for the unknown coefficients $\rho$, $T$, and $A^\pm$.

Also, for the geometry in Fig. 1 with the local thickness-dependent permittivity (7) for a WM slab (with the equivalent problem shown in Fig. 2) the transmission network (ABCD matrix) can be easily obtained,

$$\begin{bmatrix} A & B \\ C & D \end{bmatrix} = \begin{bmatrix} 1 & 0 \\ Y_{g2} & 1 \end{bmatrix} \cdot \begin{bmatrix} \cosh(\gamma_{\text{loc}} L) & Z_{\text{loc}} \sinh(\gamma_{\text{loc}} L) \\ \frac{1}{Z_{\text{loc}}} \sinh(\gamma_{\text{loc}} L) & \cosh(\gamma_{\text{loc}} L) \end{bmatrix} \begin{bmatrix} 1 & 0 \\ Y_{g1} & 1 \end{bmatrix}$$
(12)

where $Z_{\text{loc}} = \frac{\gamma_{\text{loc}}}{j\omega\varepsilon_0 \varepsilon_h}$ is the characteristic impedance of the WM slab as the transmission line with the local thickness-dependent permittivity (7). The reflection and transmission coefficients (S-parameters) are retrieved from the ABCD-matrix parameters with the known expressions from the microwave engineering [38, p. 192; with $Z_0$ replaced by $Z_{\text{loc}}$]. Obviously, in a local framework the transmission network (12) (with conversion to the S-parameters) and the field approach (11) give the same results for $\rho$ and $T$.

For a symmetric WM structure with respect to the impedance surface terminations, $Y_{g1} = Y_{g2} = Y_g$, even though the lumped impedance insertions can be different and in general $\alpha_1 \neq \alpha_2$, a simple circuit model with even and odd excitations (corresponding to the symmetry of the structure by a perfect electric conductor (PEC) and perfect magnetic conductor (PMC), respectively, placed at the center of the structure) can be obtained [39],

$$\rho_e = \frac{Y_0 - Y_g - Y_{\text{loc}} \coth\left(\gamma_{\text{loc}} \frac{L}{2}\right)}{Y_0 + Y_g + Y_{\text{loc}} \coth\left(\gamma_{\text{loc}} \frac{L}{2}\right)}$$
(13)

$$\rho_o = \frac{Y_0 - Y_g - Y_{\text{loc}} \tanh\left(\gamma_{\text{loc}} \frac{L}{2}\right)}{Y_0 + Y_g + Y_{\text{loc}} \tanh\left(\gamma_{\text{loc}} \frac{L}{2}\right)}$$
(14)

where $Y_{\text{loc}} = 1/Z_{\text{loc}}$, and $Y_0 = j\omega\varepsilon_0/\gamma_0$. The reflection/transmission response of the entire WM structure is obtained by the superposition principle as $\rho = (\rho_e + \rho_o)/2$ and $T = (\rho_e - \rho_o)/2$.

In the next section, we will present several numerical examples to validate the proposed local thickness-dependent permittivity model for far-field and near-field excitation problems. Also, we will demonstrate that the homogenized local WM can be effectively used in the full-wave numerical simulator to significantly reduce the computation time and memory in the analysis of true physical WM structures.

III. NUMERICAL RESULTS AND DISCUSSIONS

In the first numerical example we consider a mushroom-type high-impedance surface with lumped impedance insertions at the connection of the vias to the ground plane with the geometry shown in Fig. 3. The response of the structure with the TM-polarized plane-wave incidence has been studied in [37] based on the non-local homogenization model.

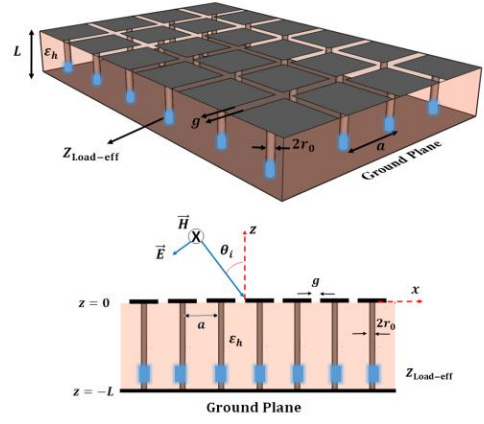

Fig. 3. Geometry of a mushroom high-impedance surface with loaded vias with an obliquely incident TM-polarized plane wave.

The local thickness-dependent permittivity model has been utilized with $\alpha_1 = C_p / C_w$ and $\alpha_2 = (j\omega C_w Z_{\text{load eff}})^{-1}$ in (7). The reflection coefficient is obtained by solving either (11) or (12), or directly using (13) (with $L/2$ replaced by $L$) corresponding to the equivalent geometry in Fig. 2. The results for the phase of the reflection coefficient with different values of inductive and capacitive loads (including the cases of open circuit (OC) and short circuit (SC) vias) are shown in Fig. 4 and compared with the exact non-local results and full-wave numerical simulations using HFSS from [37], demonstrating good agreement with the previously obtained results.

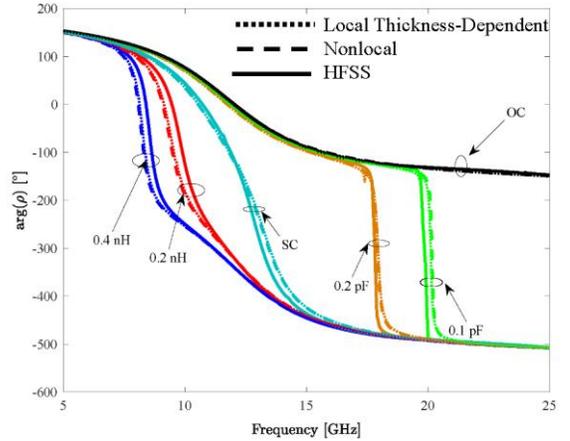

Fig. 4. Phase of the reflection coefficient for a mushroom structure with the vias connected to the ground through inductive and capacitive loads (including the cases of SC and OC). The dimensions of the structure are: $a = 2$ mm, $g = 0.2$ mm, $r_0 = 0.05$ mm, $L = 1$ mm, $\varepsilon_h = 10.2$, and $\theta_i = 60°$. The load is connected to the ground through a gap of 0.1 mm with the parasitic capacitance $C_{\text{par}} \approx 0.02$ pF and parasitic inductance $L_{\text{par}} \approx 0.06$ nH (estimated by curve fitting).

As pointed out in Section II, the local thickness-dependent permittivity (7) takes into account spatial dispersion in WM in the averaged sense per thickness of the WM slab and the effect of the loads/terminations. Fig. 5 shows the frequency dispersion of the local permittivity for a WM with the vias connected to the metallic patches at $z = 0$ and to the ground plane through the lumped inductive or capacitive loads at $z = -L$ (Fig. 3). The values of the loads and the dimensions of the structure are as in Fig. 4. It can be seen that the local permittivity depends on the type and the value of the load and resonates at the Fabry-Pérot condition associated with the effective thickness of the WM slab.

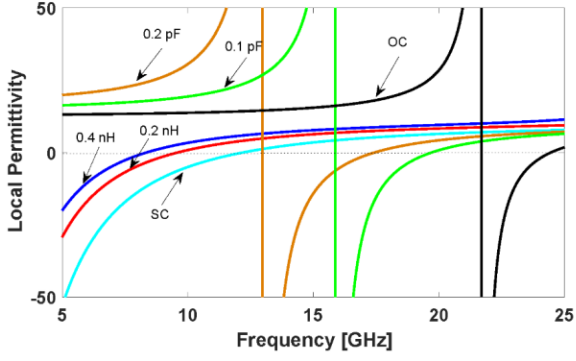

Fig. 5. Local thickness-dependent permittivity of the WM slab with the vias connected to the metallic patches and to the ground plane through the lumped loads. The various lumped loads correspond to those in Fig. 4.

In Fig. 6 the reflection phase characteristics are shown for an air-filled mushroom structure with the vias connected to inductive loads of a large value, and the results are compared with the non-local solution and the full-wave numerical simulations from [37], demonstrating nearly perfect agreement. The effects of the parasitic inductance and parasitic capacitance are negligible in such a case of an air-filled structure. It should be noted that the local thickness-dependent permittivity model accurately captures the response of the structure even with large discrete lumped loads.

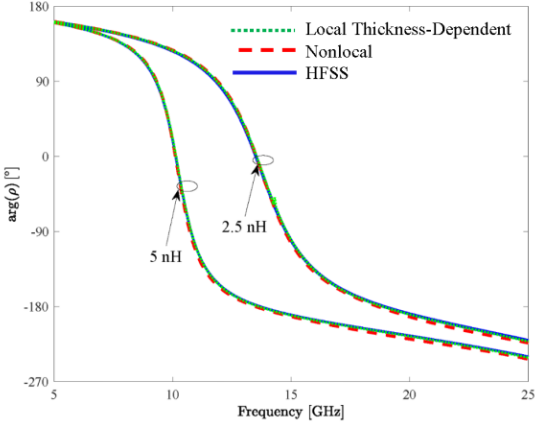

Fig. 6. Phase of the reflection coefficient for the air-filled mushroom structure with the vias connected to the ground plane through inductive loads (2.5 nH and 5 nH) (with the geometry in Fig. 3). The dimensions of the structure are: $a = 2$ mm, $g = 0.2$ mm, $r_0 = 0.05$ mm, $L = 1$ mm, $\varepsilon_h = 1$, and $\theta_i = 45°$.

In the second numerical example we consider a two-sided mushroom structure (with the same patch arrays at the WM interfaces) with the vias connected to PIN diodes in the middle of the structure (with the geometry shown in Fig. 7). This structure has been studied in [40] using a non-local homogenization model.

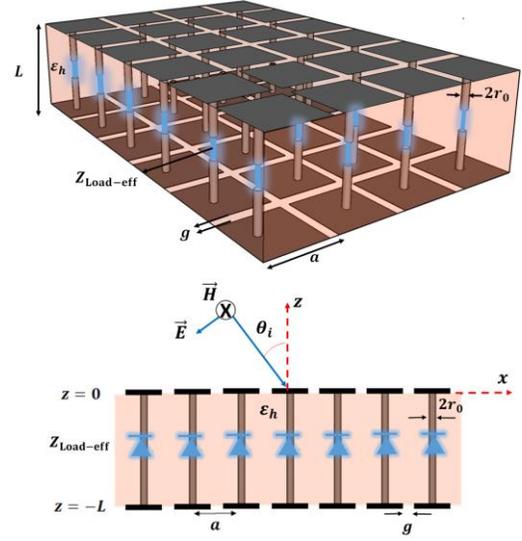

Fig. 7. Geometry of a two-sided mushroom structure with the vias connected to PIN diodes with an obliquely incident TM-polarized plane wave.

The local thickness-dependent permittivity model has been applied in the analysis of the structure due to the obliquely incident TM-polarized plane wave. Because of the symmetry of the structure, even and odd excitations can be used with $\alpha_1 = C_p / C_w$ and $\alpha_2 = \left( j\omega C_w Z_{\text{diode eff}} \right)^{-1}$ in the PEC symmetry in (7), and with $\alpha_1 = C_p / C_w$ and $\alpha_2 = 0$ in the PMC symmetry in (7). The response of the structure is obtained with (13) and (14) (for the equivalent problem in Fig. 2). The PIN diodes are modeled as effective diode loads with the impedance of diodes in the ON and OFF states as the series connection of lumped resistors and capacitors with $R = 3$ Ohms and $C = 0.025$ pF. The diodes are inserted in the vias through a gap of 0.73 mm. The parasitic loads in order to characterize the gap were estimated with the values of the parasitic capacitance of $C_{\text{par}} \approx 0.02$ pF and the parasitic inductance of $L_{\text{par}} \approx 0.1$ nH [40].

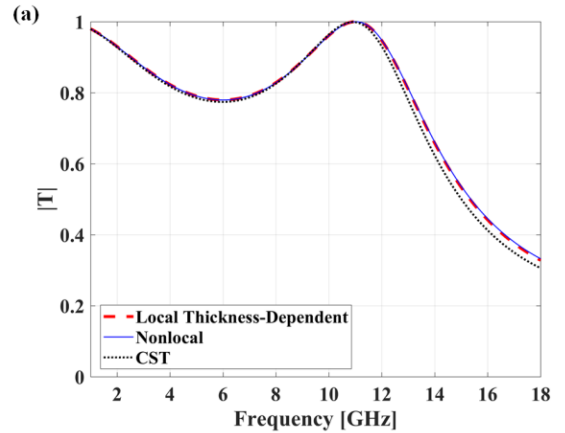

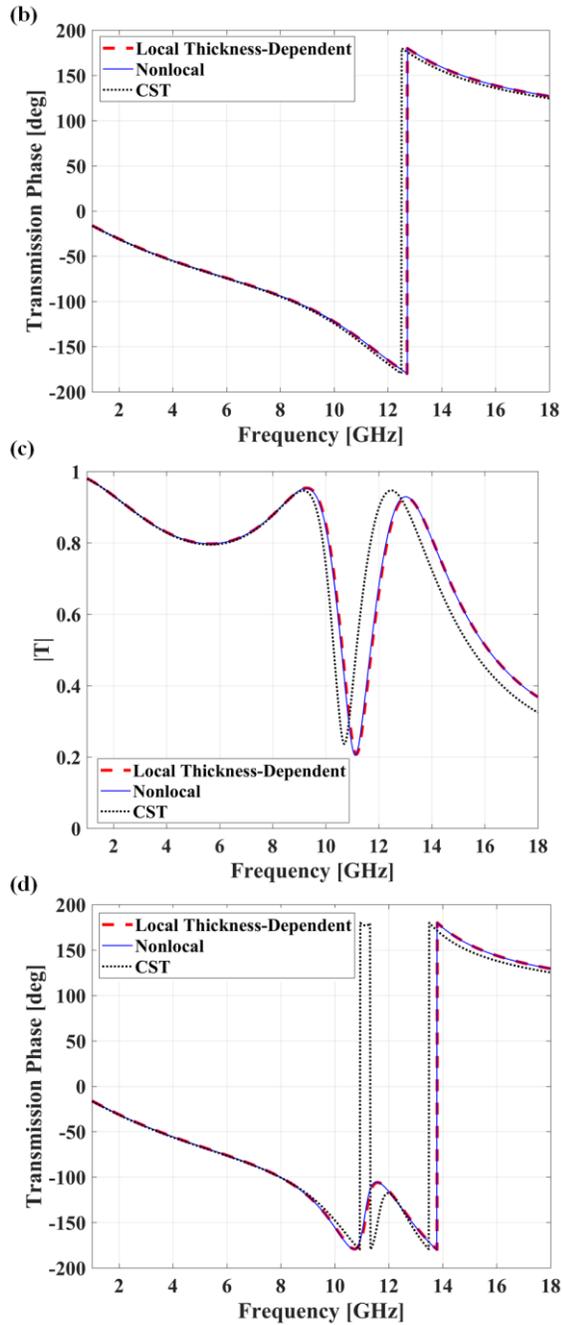

Fig. 8. Transmission response (magnitude and phase) of the two-sided mushroom structure with the vias connected to PIN diodes in the middle of the WM slab in (a)-(b) OFF and (c)-(d) ON states under illumination of an obliquely incident TM-polarized plane wave. The dimensions of the structure are: $a = 2$ mm, $g = 0.2$ mm, $r_0 = 0.05$ mm, $L = 2$ mm, $\varepsilon_h = 10.2$, and $\theta_i = 60°$.

The results of the transmission coefficient (magnitude and phase) based on the local thickness-dependent permittivity formalism are shown in Fig. 8 for both OFF (Figs. 8(a),(b)) and ON (Figs. 8(c),(d)) states and compared with the non-local and CST MWS results from [40], demonstrating great agreement.

Next, we validate the proposed local thickness-dependent permittivity model for near-field excitation, with a specific application to subwavelength imaging problems involving WM-type lenses. First, we consider a WM slab with the magnetic line source excitation (with the geometry shown in Fig. 9). The transmission coefficient for propagating and evanescent waves from the source is calculated either by (11) or (12), or with the even and odd excitation (due to symmetry of the structure) using (13) and (14) with $\alpha_1 = \alpha_2 = 0$ in (7) (see Fig. 2 for an equivalent problem). Then, the magnetic field at the image plane at a distance $d$ from the lower interface of the WM slab is calculated as a numerical solution of the Sommerfeld integral in the spectral domain [23],

$$H_y(x) = \frac{I_0 k_0^2}{j\pi\omega\mu_0} \int_0^\infty \frac{1}{2\gamma_0} e^{-\gamma_0(2d)} T(\omega, k_x) \cos(k_x x) dk_x. \quad (15)$$

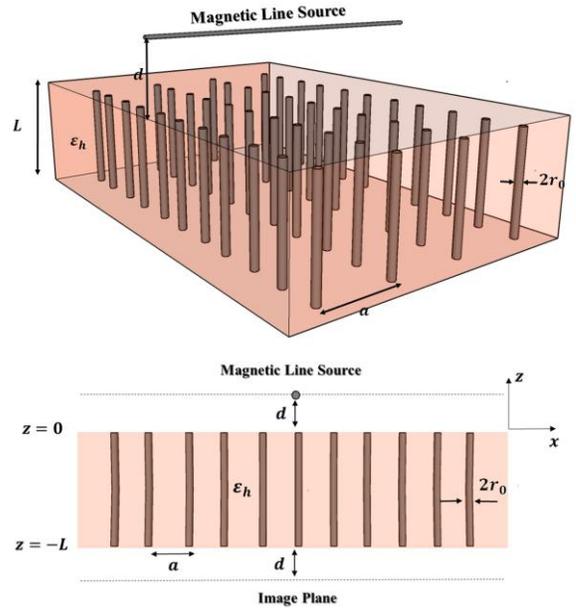

Fig. 9. Geometry of a WM slab excited by a magnetic line source.

The response of the WM slab to evanescent waves from the source is studied based on the local thickness-dependent permittivity model and compared with the non-local homogenization model results in Fig. 10 (with the non-local model formulation from [11]).

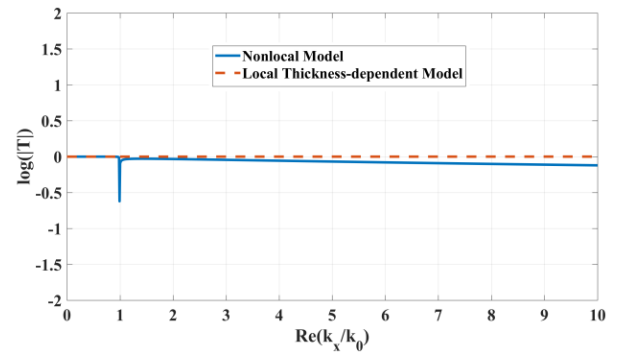

Fig. 10. Transmission response (in logarithmic units) of a WM slab for evanescent waves excited by a magnetic line source. The dimensions of the structure are: $a = 215$ nm, $r_0 = 21.5$ nm, $L = 7,894$ nm, and $\varepsilon_h = 1$.

The square normalized amplitude of the magnetic-field profile at the image plane as a function of $x/\lambda$ at the operating frequency of 19 THz is calculated as a numerical solution of (15) for both local thickness-dependent permittivity model and non-local model and shown in Fig. 11. It is assumed that the magnetic line source and the image plane are located at $d = 150$ nm. Good agreement is observed between local and non-local results.

We follow up with the example of a WM slab loaded with a 2D material (such as a graphene) at the WM interfaces and excited by a magnetic line source (with the geometry shown in Fig. 12). The non-local homogenization model and the numerical results with the line source excitation has been presented in [24]. The transmission response of the structure based on the local thickness-dependent permittivity formalism is obtained for propagating and evanescent waves from the source either by (11) or (12), or using the even and odd excitation (13) and (14) with $\alpha_1 = \alpha_2 = \sigma_s / j\omega\varepsilon_0\varepsilon_h$ in (7) (see Fig. 2 for an equivalent problem), where $\sigma_s$ is the complex surface conductivity of graphene modeled with the Kubo formula [41].

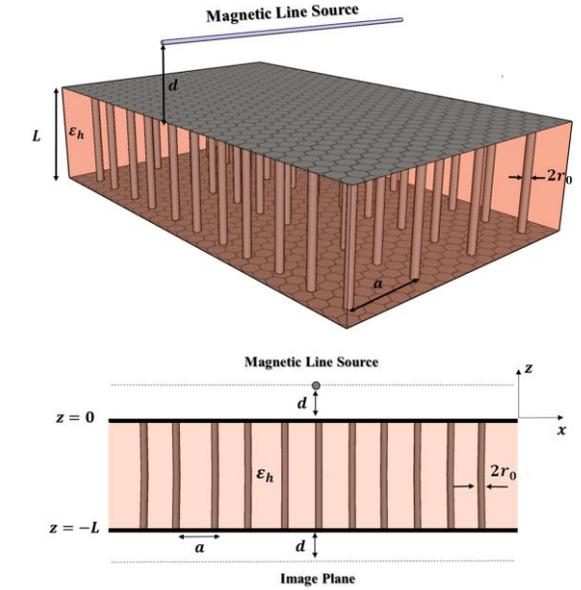

Fig. 12. Geometry of a WM slab loaded by graphene sheets at the WM interfaces and excited by a magnetic line source.

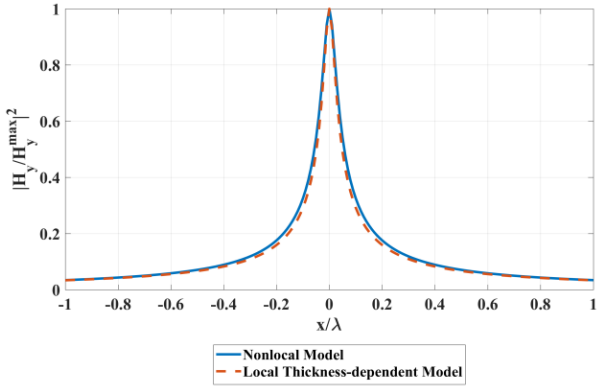

Fig. 11. Square normalized amplitude of the magnetic field at the image plane as a function of $x/\lambda$ at the operating frequency of 19 THz for a WM slab excited by a magnetic line source.

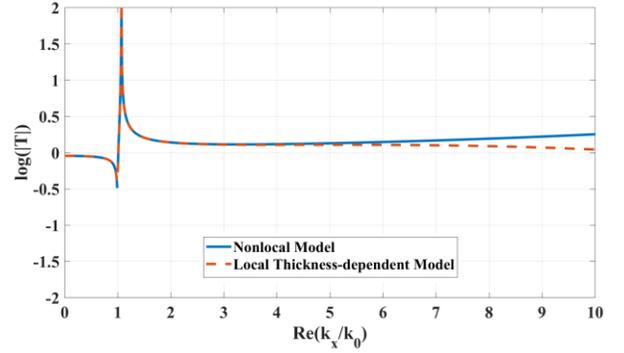

Fig. 13. Transmission response (in logarithmic units) of a WM slab loaded with graphene sheets for the evanescent waves due to a magnetic line source excitation. The dimensions of the WM slab and graphene parameters are: $a = 215$ nm, $r_0 = 21.5$ nm, $L = 2,400$ nm, $\varepsilon_h = 1$, $T = 300$ K, $\tau = 0.5$ ps, and $\mu_c = 1.5$ eV.

The response of the graphene loaded WM slab to evanescent waves is shown in Fig. 13 based on the local thickness-dependent permittivity model and compared with the non-local homogenization model results from [24], demonstrating good agreement.

The square normalized amplitude of the magnetic field at the image plane as a function of $x/\lambda$ at the operating frequency of 19 THz is shown in Fig. 14 by numerically integrating (15) with the known expression for the transmission coefficient (based local and non-local homogenization models). The location of the magnetic line source and the image plane is the same as in the previous example, $d = 150$ nm. The results of both analytical models are in very good agreement.

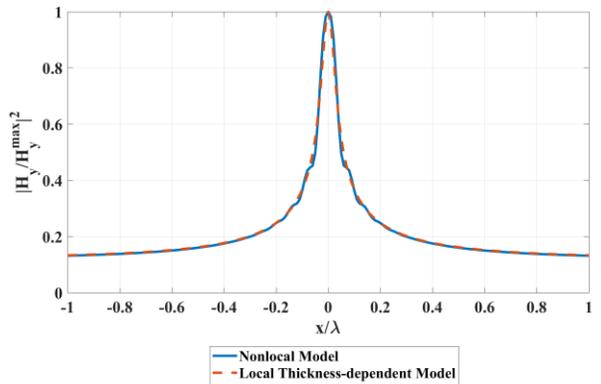

Fig. 14. Square normalized amplitude of the magnetic field at the image plane as a function of $x/\lambda$ at the operating frequency of 19 THz for a WM slab loaded with graphene sheets at the WM interfaces and excited by a magnetic line source.

In the final examples, the obtained analytical expression (7)

for the local thickness-dependent permittivity of WM has been used in the full-wave numerical simulator CST Microwave Studio [42]. The WM slab is modeled in CST MWS as a uniform local anisotropic material with the host permittivity in the $x-$ and $y-$ directions and the local thickness-dependent permittivity along the $z-$ direction (see Fig. 2 for an equivalent problem). This significantly reduces the computation time and memory when modeling true physical WM structures. The examples of a WM slab and WM slab loaded with graphene sheets at the interfaces and excited by a magnetic line source considered above are included here to demonstrate the idea of using a homogenized material in the full-wave simulator. The same dimensions are used for WM and parameters of graphene as in the previous examples (Fig. 9 and Fig. 12). In Fig. 15 for the case of a WM slab excited by a magnetic line source (Fig. 9) the magnetic-field distribution is shown for homogenized and true physical structures, with the results for the square normalized amplitude of the magnetic field at the image plane demonstrated in Fig. 16.

The same accuracy of 250 mesh cells are used in both simulations for homogenized and true physical WM slab. Table I documents the simulation time and the number of mesh cells used in the CST MWS simulations.

TABLE I

Comparison of computation time and memory in CST simulations for a WM slab excited by a magnetic line source

| Structure | Time (minutes) | Mesh Cells (millions) |
|---|---|---|
| True WM slab | 35 | 5.703 |
| Local permittivity model | 8 | 1.139 |

The numerical simulation results for the magnetic-field distribution for the example of a WM slab loaded with graphene sheets at the interfaces (Fig. 12) are shown in Fig. 17 for homogenized and true physical structures.

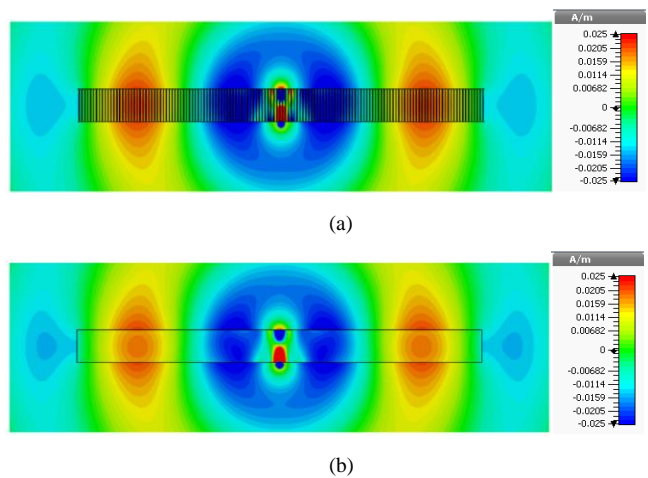

(a)

(b)

Fig. 17. Numerical simulations of the magnetic-field distribution for a (a) true physical WM slab loaded with graphene sheets and (b) homogenized local anisotropic material loaded with graphene sheets and excited by a magnetic line source.

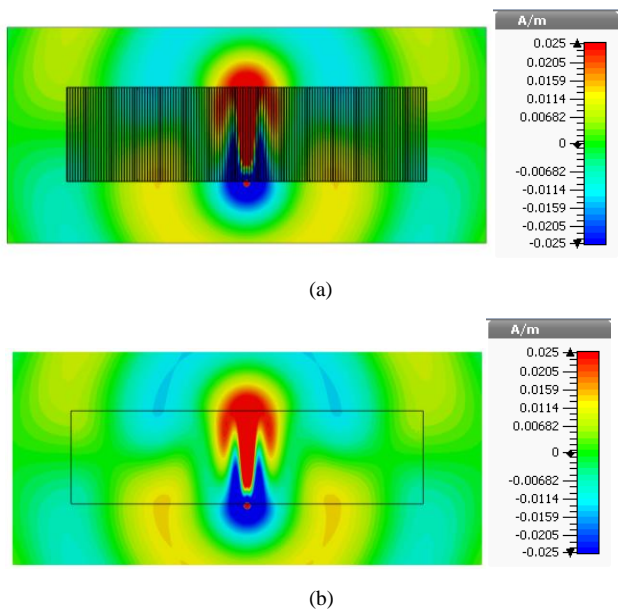

(a)

(b)

Fig. 15. Numerical simulations of the magnetic-field distribution for a (a) true physical WM slab and (b) homogenized local anisotropic material excited by a magnetic line source.

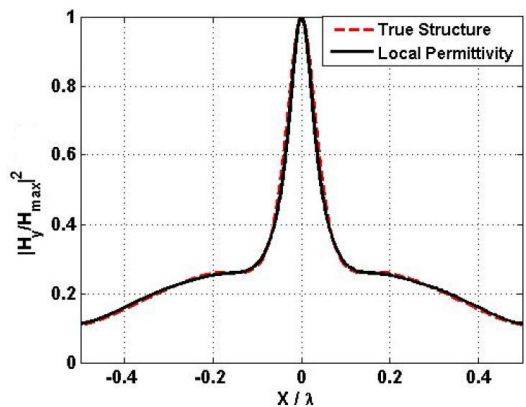

Fig. 18. Square normalized amplitude of the magnetic field at the image plane for a true physical WM slab with graphene sheets and homogenized local anisotropic slab with graphene sheets excited by a magnetic line source.

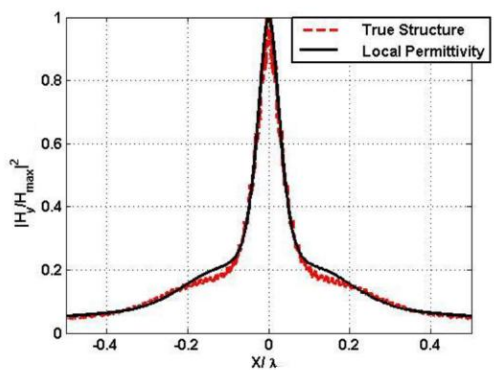

Fig. 16. Square normalized amplitude of the magnetic field at the image plane for a true physical WM slab and homogenized local anisotropic slab excited by a magnetic line source.

The simulation results for the square normalized amplitude of the magnetic field at the image plane are shown in Fig. 18. The same accuracy of 200 mesh cells are used in both simulations for homogenized and true physical structures. Table II documents the simulation time and the number of mesh cells used in the CST MWS simulations.

TABLE II

Comparison of computation time and memory in CST simulations for a WM slab loaded with graphene sheets and excited by a magnetic line source

| Structure | Time (minutes) | Mesh Cells (millions) |
|---|---|---|
| True WM slab | 38 | 7.604 |
| Local permittivity model | 17 | 3.245 |

IV. CONCLUSION

A local thickness-dependent permittivity model has been proposed for a general case of a WM structure with the wires connected to impedance surfaces through lumped impedance insertions. A closed-form expression of a local thickness-dependent permittivity has been derived which takes into account spatial dispersion in WM in the average sense per thickness of the WM slab and the effect of lumped loads and impedance surface terminations. The local model necessitates the solution for the non-local permittivity in the spatial domain, however, once it is derived, it enables to obtain a general closed-form expression for the local thickness-dependent permittivity, which simplifies the formulation for various geometrically-complex excitation and scattering problems involving WM. In the local framework the WM slab with lumped impedance insertions and terminated with impedance surfaces is replaced by a local uniaxial anisotropic material subject to traditional boundary conditions at the interfaces (without the need for ABCs required in the non-local model). It is demonstrated with various numerical examples that the local permittivity framework provides accurate solutions for far-field and near-field electromagnetic problems. In general, the proposed theory captures accurately the non-local wave dynamics of the WM for metamaterial slabs with a thickness roughly less than 0.8 of the wavelength in the host medium [30]. We have also introduced the local thickness-dependent permittivity in the full-wave numerical simulator CST Microwave Studio and have demonstrated a significant reduction of computation time and memory when modeling true physical WM structures as homogenized anisotropic materials.

Moreover, the theory highlights that the dependence of the effective permittivity of "local" models on the material thickness, which is a feature common to most metamaterials, is a consequence of a nonlocal electromagnetic response.

V. ACKNOWLEDGEMENT

The authors would like to thank Ali Forouzmand for his help regarding the validation of local thickness-dependent permittivity model for near-field excitation and Gabriel Moreno for helping with CST Microwave Studio simulations.

APPENDIX

GREEN'S FUNCTION PROBLEM

For a general case of a WM structure shown in Fig. 1 the Green's function problem is formulated for the wave equation (3) subject to the ABCs at the wire-end connections at $z=0$ and $z=-L$ [25]-[29],

$$\left(\frac{\partial^2}{\partial z^2}+k_h^2\right)g(z,z')=-\delta(z-z') \quad (16)$$

$$\left[g(z,z')+\alpha_1\frac{\partial g(z,z')}{\partial z}\right]\Big|_{z=0}=0 \quad (17)$$

$$\left[g(z,z')-\alpha_2\frac{\partial g(z,z')}{\partial z}\right]\Big|_{z=-L}=0. \quad (18)$$

The solution of the boundary-value problem (16)-(18) is obtained as follows,

$$g(z,z')=\frac{e^{-jk_h|z-z'|}}{2jk_h}-\frac{e^{-jk_h(z-z')}}{2jk_h}+\left(e^{jk_hz}-e^{-jk_hz}\frac{1+jk_h\alpha_1}{1-jk_h\alpha_1}\right)B(z') \quad (19)$$

where

$$B(z')=\frac{1-jk_h\alpha_1}{2jk_h}$$

$$\times\frac{e^{jk_h(z'+L)}(1+jk_h\alpha_2)-e^{-jk_h(z'+L)}(1-jk_h\alpha_2)}{e^{-jk_hL}(1-jk_h\alpha_1)(1-jk_h\alpha_2)-e^{jk_hL}(1+jk_h\alpha_1)(1+jk_h\alpha_2)}. \quad (20)$$

In (19), (20), $\alpha_1$ and $\alpha_2$ depend on the lumped impedance insertions and impedance surface terminations, such that for lumped loads attached to metallic patches,

$$\alpha_1=\left(j\omega C_w Z_{\text{load eff1}}+\frac{C_w}{C_{p1}}\right)^{-1},\quad \alpha_2=\left(j\omega C_w Z_{\text{load eff2}}+\frac{C_w}{C_{p2}}\right)^{-1} \quad (21)$$

and for lumped loads connected to thin metal/2D material,

$$\alpha_1=\left(j\omega C_w Z_{\text{load eff1}}+\frac{j\omega\varepsilon_0\varepsilon_h}{\sigma_{s1}}\right)^{-1},\quad \alpha_2=\left(j\omega C_w Z_{\text{load eff2}}+\frac{j\omega\varepsilon_0\varepsilon_h}{\sigma_{s2}}\right)^{-1}. \quad (22)$$

Here, $Z_{\text{load eff1,2}}$ are defined in [23], [28], [37], $C_w$ and $C_{p1,2}$ are given in [43], and $\sigma_{s1,2}$ determines a complex surface conductivity of 2D material.